\documentclass{article}
\usepackage{graphicx}

\newcommand{\rd}{\mathrm{d}}

\begin{document}

\title{Euler's elastica in nonlocal theory of elasticity}
\author{Vasyl Kovalchuk, Ewa Eliza Ro\.zko, Barbara Go\l{}ubowska}
\date{}
\maketitle

\begin{abstract}
A generalization of the Euler's elastic problem, i.e., finding a stationary configuration (planar elastica) of the Bernoulli's thin ideal elastic rod with boundary conditions defined through fixed endpoints and/or tangents at the endpoints, for the chosen nonlocal differential constitutive stress-strain relation (i.e., nonlocal theory of elasticity) is considered. In the classical (local) Euler-Bernoulli's beam model, the general solutions of the governing equations (that are inhomogeneous but linear) for bending moments and shear forces in the case of large deformations can be obtained using the Jacobi elliptic functions and incomplete elliptic integrals. For the discussed nonlocal toy differential model, the general solutions of the governing equations (that are this time nonlinear) can also be expressed in the parametric form through the linear combinations of all three incomplete elliptic integrals. As further research, we plan to apply some boundary conditions (clamped, simply supported, etc.) for the obtained nonlocal general solutions in order to compare them to the local solutions for the corresponding boundary value problems.\\[0.2cm]
\end{abstract}

\section{Introduction}\label{sec:1}

The classical Euler-Bernoulli's model was originally developed in the XVIII-th century in order to describe the large deformations of plane curved beams based on the local elasticity theory. It can be shown that the exact solution of this problem can be written using the Jacobi elliptic functions and incomplete elliptic integrals of the first, second, and third kind (see, e.g., the paper \cite{bib3} in application to the analysis of the lightweight shock absorbing structures formed from many arc-curved beams placed between two flat platforms, where the curved beams can store more energy and produce less reaction forces compared to the ordinary elastic structures).

The Euler-Bernoulli's theory can be applied also to very small objects, e.g., to nanobeams (including nanowires \cite{bib8}, nanotubes, nanorods, etc.), considered as beams with small length scale (nanoscale) that can be deformed with bending moments, as well as shear and axial forces. Such objects in nanoscale exhibit extraordinary physical and mechanical characteristics: high aspect ratio, high flexibility, high tensile and shear strength, and high modulus of elasticity. For instance, for carbon nanotubes we have that \cite{bib6}
\begin{eqnarray}
&&\rho = 2300\ \frac{{\rm kg}}{{\rm m}^{3}},\quad  E = 1000\ {\rm  GPa},\quad \nu = 0.19,\quad G = 420\ {\rm GPa}\nonumber\\
\label{eq1.1-2}\\
&&d = 1.0\ {\rm nm},\ A = 0.785\ {\rm nm}^{2},\ I=\frac{\pi d^{4}}{64} = 0.0491\ {\rm nm}^{4},\ l_{i} = 1.5\ {\rm nm}\nonumber
\end{eqnarray}
where \(\rho\) is the density, \(E\) is the Young's (elastic) modulus, \(\nu\) is the Poisson's ratio, \(G\) is the Kirchhoff's (shear) modulus, \(d\) is the diameter of nanotubes, \(A\) is their section area, \(I\) is the moment of inertia of their section, and \(l_{i}\) is the internal characteristic length.

In order to describe nanobeams not only the classical beam theory is used, but also the nonlocal elasticity (in relation to the small-scale effects) is applied that allows us to investigate, for instance, the problems of static bending, free vibration analysis, and also elastic buckling of carbon nanotubes (see, e.g., the papers \cite{bib5,bib6}, as well as \cite{bib7} where the postbuckling of unknown-length nanobeams, that is based on the concept of variable-arc-length (VAL) beams, is investigated).

\section{Explicit analytical solutions for local elastica}\label{sec:2}

The formal mathematical description of the elastic line or Euler's elastica can be found, e.g., in the papers \cite{bib1,bib2} or in the paper \cite{bib4}, where the general solutions in terms of elliptic functions and explicit parametrizations for the free elastica as well as the elastica with tension are presented.

The equilibrium equations of the small segment of the beam in the theory of elasticity are given as
\begin{equation}\label{eq2.1}
\frac{\rd M}{\rd s} = Q,\qquad \frac{\rd Q}{\rd s} = - \kappa N,\qquad \frac{\rd N}{\rd s} = \kappa Q
\end{equation}
where \(M\) is the bending moment, \(N\) and \(Q\) are the axial and shear forces, and \(\kappa\) is the curvature of the elastica that is parametrized by the so-called arc length parameter \(s\) changing from 0 to \(L\), where \(L\) is the total length of the elastica (sometimes, without loss of generality, the unit elastica is considered for which \(L = 1\)).

If we define \(\theta(s)\) as the tangent slope angle at any point \(P(x(s),y(s))\) of the elastica that is situated along the \(x\)-axis and subjected to the compressive load \(F\) also directed along the \(x\)-axis, then the buckling (e.g., the transversal deflection) of the
Euler's elastica will happen along the \(y\)-axis. In such a situation the axial and shear forces, the curvature, and the geometrical conditions will be given as
\begin{equation}\label{eq2.2}
N = - F\cos\theta,\quad Q = F\sin\theta,\quad \frac{\rd\theta}{\rd s} = \kappa,\quad \frac{\rd x}{\rd s} = \cos\theta,\quad \frac{\rd y}{\rd s} = \sin\theta.
\end{equation}

In the classical (local) beam model the bending moment at any point of the elastica is proportional to its curvature, i.e., the bending moment-curvature relation is governed by the Euler-Bernoulli's law as
\begin{equation}\label{eq2.3}
M = - EI\kappa = - EI\frac{\rd\theta}{\rd s}
\end{equation}
with the flexural rigidity \(EI\) being expressed through the Young's modulus \(E\) and the moment of inertia \(I\).

Additionally, it can be shown \cite{bib2} that the bending moment \(M\) and the axial force \(N\) acting along the elastic curve are related through the equation
\begin{equation}\label{eq2.4}
\frac{M^{2}}{2EI} = |N|
\end{equation}
that can be understood as the manifestation of the demand that in the state of mechanical equilibrium we have that the sum of forces at all point of the elastica should be zero.

Using (\ref{eq2.1})--(\ref{eq2.3}) we can obtain straightforwardly the governing equation as
\begin{equation}\label{eq2.5}
\frac{\rd^{2}\theta}{\rd s^{2}} = - \frac{1}{EI}\frac{\rd M}{\rd s} = - \frac{Q}{EI} = - \frac{F}{EI}\sin\theta = - \alpha^{2}\sin\theta,\qquad \alpha^{2} = \frac{F}{EI}\cdot\ \
\end{equation}
The above inhomogeneous linear second-order ordinary differential equation on the function \(\theta(s)\) can be once integrated when we multiply the left- and right-hand sides of (7) by the term \(2(\rd\theta/\rd s)\). Then we will obtain that
\begin{equation}\label{eq2.6}
\frac{\rd}{\rd s}\bigl(\bigl(\frac{\rd\theta}{\rd s}\bigr)^{2}\bigr)=2\alpha^{2}\frac{\rd}{\rd s}\left(\cos\theta\right)
\end{equation}
can be integrated as
\begin{equation}\label{eq2.7}
\frac{\rd\theta}{\rd s} =\pm \sqrt{2\alpha^{2}\cos\theta + C} = \pm\frac{2\alpha}{k}\sqrt{1 - k^{2}\sin^{2}\frac{\theta}{2}},\qquad k^{2} = \frac{4\alpha^{2}}{C + 2\alpha^{2}}
\end{equation}
where \(C\), or equivalently \(k\) (if \(\alpha\neq 0\), i.e., \(F\neq 0\)), is the first integration constant. It is worth to notice that the integration constant $C$ is not dimensionless, but has a dimensionality nm$^{-2}$, i.e., the same dimensionality as that of the term $\alpha^2$.

In the case when \(k^{2} \leq 1\) (i.e., $C\geq 2\alpha^2$), the above expression can be integrated as
\begin{equation}\label{eq2.8}
s(\theta) =\pm\frac{k}{2\alpha}\int\frac{\rd \theta}{\sqrt{1 - k^{2}\sin^{2}\frac{\theta}{2}}} = s_0\pm\frac{k}{\alpha}F\bigl(\frac{\theta}{2},k\bigr)
\end{equation}
where $s_0$ is the second integration constant and
\begin{equation}\label{eq2.9}
F(\varphi,k) = \int_{0}^{\varphi}\frac{\rd\varphi}{\sqrt{1 - k^{2}\sin^{2}\varphi}}
\end{equation}
is the incomplete elliptic integral of the first kind.

Thus, inverting the expression (\ref{eq2.8}), i.e., noticing that the Jacobian elliptic sine function ${\rm sn}(u,k)=\sin\varphi$ is a simple inverse of the incomplete elliptic integral of the first kind \(F(\varphi,k)=u\), the general solution of (\ref{eq2.5}) can be
written as \cite{bib3}
\begin{equation} \label{eq2.10ab}
\theta(s)=\pm 2\arcsin\bigl({\rm sn}\bigl(\frac{\alpha}{k}(s-s_0),k \bigr) \bigr),  \qquad
\frac{\rd\theta}{\rd s}=\pm\frac{2\alpha}{k}{\rm dn}\bigl( \frac{\alpha}{k}(s-s_0),k \bigr)
\end{equation}
where we have used the connection between the Jacobi elliptic delta and sine functions for rewriting (\ref{eq2.7}) in the form of (\ref{eq2.10ab}), i.e.,
\begin{equation}\label{eq2.10c}
{\rm dn}^2(u,k)=1-k^2{\rm sn}^2(u,k)=1-k^2\sin^2\varphi.
\end{equation}

Both integration constants, i.e., $k$ (or $C$) and $s_0$, can be defined through the application of the corresponding boundary conditions (clamped, simply supported, etc.) for the boundary value problems (BVPs) of the curved beam.

Using (\ref{eq2.10ab}) we can obtain that
\begin{eqnarray}
\sin\theta&=&\pm 2\;{\rm sn}\bigl( \frac{\alpha}{k}(s-s_0),k \bigr){\rm cn}\bigl( \frac{\alpha}{k}(s-s_0),k \bigr)\nonumber\\
\label{eq2.11ab}\\
\cos\theta&=&1 - 2\;{\rm sn}^{2}\bigl( \frac{\alpha}{k}(s-s_0),k \bigr)\nonumber
\end{eqnarray}
where we have used the connection between the Jacobi elliptic cosine and sine functions, i.e.,
\begin{equation}\label{eq2.11c}
{\rm cn}^2(u,k)=1-{\rm sn}^2(u,k)=1-\sin^2\varphi=\cos^2\varphi.
\end{equation}

\begin{figure}[t]
\centering
\includegraphics[width=9.5cm,keepaspectratio]{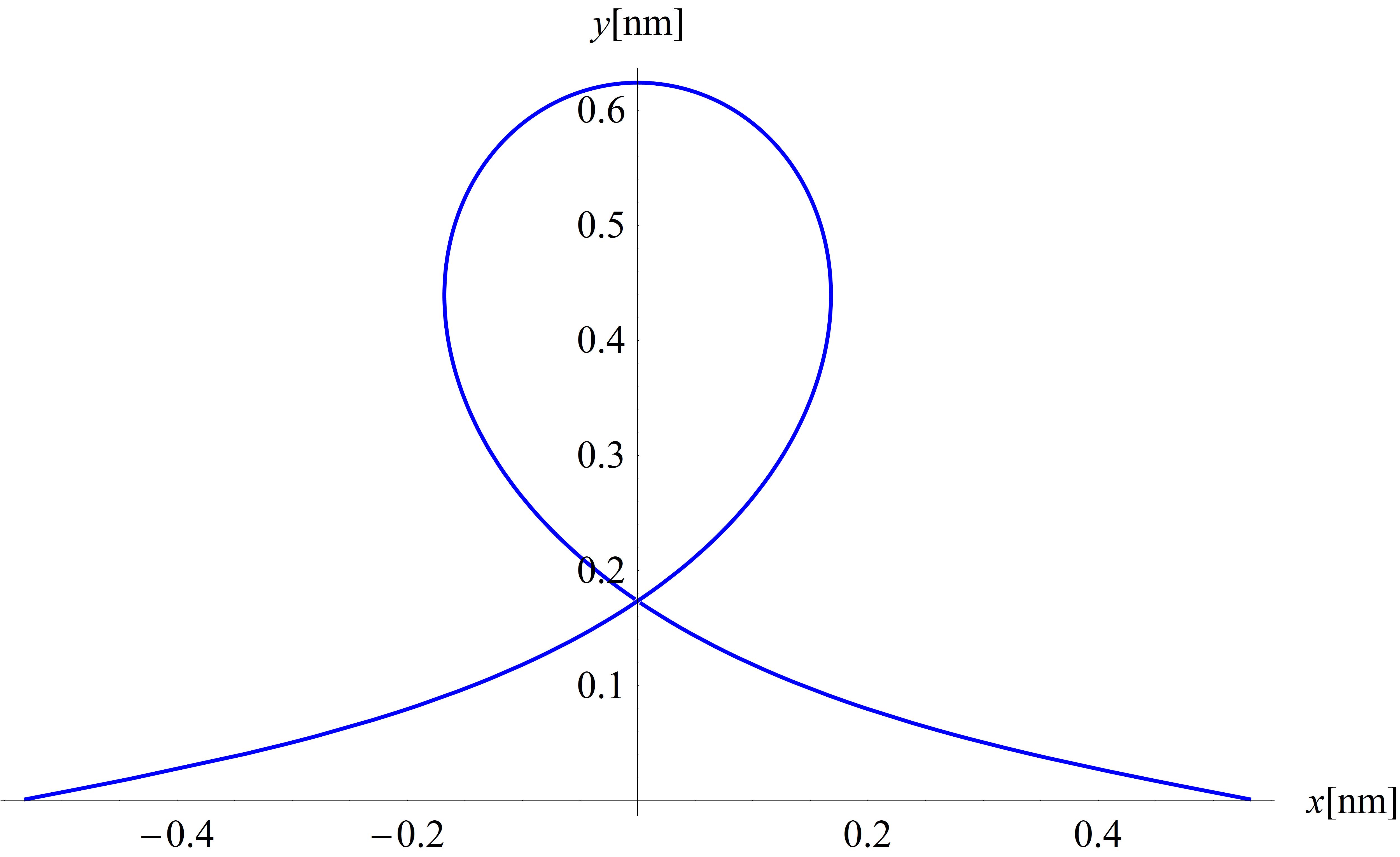}
\caption{Exemplary graph of the parametric solution $(x\left(\theta\right),y\left(\theta\right))$ defined by (\ref{eq2.12a}) and (\ref{eq2.12b}) for integration constants $x_0=y_0=0$, $C=20$ nm$^{-2}$, elliptic modulus $1/k=0.99549$, force $F=500$ nN, Young modulus $E=1000$ GPa, moment of inertia $I=0.0491$ nm$^{4}$.} \label{fig-1}
\end{figure}

Next, integrating the last two equations from (\ref{eq2.2}) and using (\ref{eq2.7}), we obtain the coordinates of any point on the Euler's elastica (see Fig.~\ref{fig-1} for the exemplary graph of the obtained general solution), i.e.,
\begin{eqnarray}
x(\theta)&=&\int{\cos\theta}\,\rd s(\theta)=\pm\frac{k}{2\alpha}\int\frac{1-2\sin^2\frac{\theta}{2}}{\sqrt{1 - k^{2}\sin^{2}\frac{\theta}{2}}}\;\rd \theta\nonumber\\
\label{eq2.12a}\\
&=&x_0\pm\frac{k}{\alpha}\bigl(1-\frac{2}{k^{2}}\bigr)F\bigl(\frac{\theta}{2},k\bigr)
\pm\frac{2}{\alpha k}E\bigl(\frac{\theta}{2},k\bigr)\nonumber\\
\nonumber\\
y(\theta)&=&\int{\sin\theta}\,\rd s(\theta)=\mp\frac{1}{2\alpha^2}\int\frac{\rd(2\alpha^2\cos\theta+C)}{\sqrt{2\alpha^{2}\cos\theta+C}}\nonumber\\
\label{eq2.12b}\\
&=&y_0\mp\frac{1}{\alpha^2}\sqrt{2\alpha^{2}\cos\theta + C}=y_0\mp\frac{2}{\alpha k}\sqrt{1 - k^{2}\sin^{2}\frac{\theta}{2}}\nonumber
\end{eqnarray}
where $x_0$, $y_0$ are integration constants and
\begin{equation}\label{eq2.12c}
E(\varphi,k) = \int_{0}^{\varphi}{\sqrt{1 - k^{2}\sin^{2}\varphi}\ \rd\varphi}
\end{equation}
is the incomplete elliptic integral of the second kind.

Using (\ref{eq2.10ab}) the expressions (\ref{eq2.12a})--(\ref{eq2.12b}) can be rewritten also as functions of $s$, i.e.,
\begin{eqnarray}
x(s)&=&x_0+\bigl(1-\frac{2}{k^{2}}\bigr)(s-s_0)
\pm\frac{2}{\alpha k}E\bigl({\rm am}\bigl( \frac{\alpha}{k}(s-s_0),k\bigr),k\bigr)\nonumber\\
\label{eq2.12de}\\
y(s)&=&y_0\mp\frac{2}{\alpha k}{\rm dn}\bigl(\frac{\alpha}{k}(s-s_0),k\bigr)\nonumber
\end{eqnarray}
where ${\rm am}(u,k)=\varphi$ is the Jacobi elliptic amplitude.

Similarly, in the case when $k^{2} > 1$ (i.e., $|C|< 2\alpha^2$), using the transformation
\begin{equation}\label{eq2.13b}
F\left(\varphi,k\right)=\frac{1}{k}\;F\bigl(\arcsin\left(k\sin\varphi\right),\frac{1}{k}\bigr)
\end{equation}
we obtain that the general solution (\ref{eq2.8}) can be rewritten as
\begin{equation}\label{eq2.13a}
s(\theta)=s_0\pm\frac{1}{\alpha}\;F\bigl(\arcsin\bigl(k\sin\frac{\theta}{2}\bigr),\frac{1}{k}\bigr).
\end{equation}
Inverting the expression (\ref{eq2.13a}) we obtain that \cite{bib3}
\begin{equation}\label{eq2.13cd}
\theta(s)=\pm 2\arcsin\bigl(\frac{1}{k}\;{\rm sn}\bigl(\alpha(s-s_0),\frac{1}{k} \bigr) \bigr),\ \ \frac{\rd\theta}{\rd s}=\pm \frac{2\alpha}{k}{\rm cn}\bigl( \alpha(s-s_0),\frac{1}{k} \bigr)
\end{equation}
therefore,
\begin{eqnarray}
\sin\theta &=&\pm\frac{2}{k}\;{\rm sn}\bigl( \alpha(s-s_0),\frac{1}{k} \bigr){\rm dn}\bigl( \alpha(s-s_0),\frac{1}{k} \bigr)\nonumber\\
\label{eq2.14ab}\\
\cos\theta&=&1 - \frac{2}{k^{2}}\;{\rm sn}^{2}\bigl( \alpha(s-s_0),\frac{1}{k} \bigr).\nonumber
\end{eqnarray}

Finally, integrating the last two equations from (\ref{eq2.2}) and using (\ref{eq2.7}), we obtain the coordinates of any point on the Euler's elastica representing the deformed beam as
\begin{eqnarray}
x(s)&=&\int{\cos\theta}\,\rd s=s-s_0-\frac{2}{k^{2}}\int{\rm sn}^{2}\bigl(\alpha(s-s_0),\frac{1}{k}\bigr)\rd s\nonumber\\
\label{eq2.15}\\
&=&x_0-s+s_0\pm\frac{2}{\alpha}E\bigl({\rm am}\bigl( \alpha(s-s_0),\frac{1}{k} \bigr),\frac{1}{k}\bigr)\nonumber\\
\nonumber\\
y(s)&=&\int{\sin\theta}\,\rd s=\pm\frac{2}{k}\int{\rm sn}\bigl( \alpha(s-s_0),\frac{1}{k} \bigr){\rm dn}\bigl( \alpha(s-s_0),\frac{1}{k} \bigr)\rd s
\nonumber\\
\label{eq2.16}\\
&=&y_0\mp\frac{2}{\alpha k}{\rm cn}\bigl(\alpha(s-s_0),\frac{1}{k}\bigr).\nonumber
\end{eqnarray}

\section{Toy differential model for nonlocal elastica}\label{sec:3}

The small-scale effects (e.g., nonlocality) can be combined with the classical (local) theory of elasticity using the corresponding modifications (in the integral or differential forms) of the constitutive relations between the normal stress and strain of the beams or nanobeams. According to the Eringen's approach (see, e.g., \cite{bib9}), we can introduce the so-called stain-driven nonlocality into the linear theory of elasticity through the integral form of the general stress-strain relation given as
\begin{equation}\label{eq3.1a1}
\sigma^{nl}_{ij}\bigl(\vec{x}\bigr)=\int k_{\mu}\bigl(|\vec{x}-\vec{x}{\:}^{\prime}|\bigr)\sigma^{l}_{ij}\bigl(\vec{x}{\:}^{\prime}\bigr)\rd V^{\prime}=E\int k_{\mu}\bigl(|\vec{x}-\vec{x}{\:}^{\prime}|\bigr)\varepsilon_{ij}\bigl(\vec{x}{\:}^{\prime}\bigr)\rd V^{\prime}
\end{equation}
where $\sigma^{nl}_{ij}\bigl(\vec{x}\bigr)$ denotes the nonlocal stress at point $\vec{x}$, $\sigma^{l}_{ij}\bigl(\vec{x}{\:}^{\prime}\bigr)=E\varepsilon_{ij}\bigl(\vec{x}{\:}^{\prime}\bigr)$ defines the local (and linear) stress-strain relation at reference point $\vec{x}{\:}^{\prime}$, while $k_{\mu}\bigl(|\vec{x}-\vec{x}{\:}^{\prime}|\bigr)$
is a decaying function that represents the long-range interactions in the beams or nanobeams. This kernel function depends on the scaling factor \(\mu = \epsilon_{0}^{2}l_{i}^{2}\) understood as the parameter describing the degree of nonlocality (with the dimensionality of the squared length). In a general case, it is defined as a function of the material parameter \(\epsilon_{0}\) and internal characteristic length \(l_{i}\) (understood, e.g., as the lattice parameter, the granular size, the distance between C-C bonds, etc.) \cite{bib6,bib7}.

In Eringen's nonlocal theory we assume that the above nonlocal kernel $k_{\mu}\bigl(|\vec{x}-\vec{x}{\:}^{\prime}|\bigr)$ is the Green's function of a specific linear differential operator $\mathcal{L}_{\mu}$ such that 
\begin{equation}
\mathcal{L_{\mu}}k_{\mu}\bigl(|\vec{x}-\vec{x}{\:}^{\prime}|\bigr)=\delta\bigl(|\vec{x}-\vec{x}{\:}^{\prime}|\bigr).
\end{equation}

The most common choice of the above differential operator is $\mathcal{L}_{\mu}=1-\mu\nabla^2$, where $\nabla^2$ is the Laplacian. This means that we can equivalently rewrite the integral form of the stress-strain constitutive relation given by (\ref{eq3.1a1}) in the corresponding differential form (see, e.g., \cite{bib6,bib7}):
\begin{equation}
\mathcal{L}_{\mu}\sigma^{nl}_{ij}=\sigma^{nl}_{ij}-\mu\nabla^2\sigma^{nl}_{ij}=\sigma^{l}_{ij}=E\varepsilon_{ij}.
\end{equation} 

In our paper we consider a toy differential model for description of nonlocal Euler's elasticas, where instead of the Laplacian with respect to the spatial variable $\vec{x}(s)$ (parametrized by the arc length parameter $s$), i.e., $\nabla^2_{\vec{x}(s)}$, we use the Laplacian with respect to the material variable $s$ (the arc length parameter itself), i.e., $\nabla^2_{s}$.

Then, in our case, the nonlocal stress-strain relation has the form
\begin{equation}\label{eq3.1}
\sigma_{xx} - \mu\frac{\rd^{2}\sigma_{xx}}{\rd s^{2}} = E\varepsilon_{xx}
\end{equation}
where the normal (along the \(x\)-axis) nonlocal stress \(\sigma_{xx}\) calculated at some point \(P(s)\) with the arc length parameter \(s\) depends not only on the normal strain \(\varepsilon_{xx}\) at this point \(P(s)\), but also on the normal strain's values at all other points of the Euler's elastica (in other words, we have the so-called strain-driven nonlocality).

The above nonlocal constitutive relation (\ref{eq3.1}) applied to the Euler-Bernoulli's beam theory leads us to modification of the bending moment-curvature relation (\ref{eq2.3}) that for our toy differential model can be now expressed as
\begin{equation}\label{eq3.2}
M - \mu\frac{\rd^{2}M}{\rd s^{2}} = - EI\kappa = - EI\frac{\rd\theta}{\rd s}\cdot
\end{equation}

Then from (\ref{eq2.1})-(\ref{eq2.2}) we can derive that
\begin{equation}\label{eq3.3}
\frac{\rd^{2}M}{\rd s^{2}} = \frac{\rd Q}{\rd s} = - \kappa N = F\cos\theta\frac{\rd\theta}{\rd s}\cdot
\end{equation}
Substituting (\ref{eq3.3}) into (\ref{eq3.2}) we will obtain that
\begin{equation}\label{eq3.4}
M = - EI\left( 1 - \mu\alpha^{2}\cos\theta \right)\frac{\rd\theta}{\rd s},\qquad \alpha^{2} = \frac{F}{EI}
\end{equation}
which can be reduced to (5) when the nonlocality parameter \(\mu\) approaches 0.

Differentiating (\ref{eq3.4}) with respect to the arc length parameter \(s\) and using (\ref{eq2.1})-(\ref{eq2.2}) again, we obtain the modification of the governing equation (\ref{eq2.5}) given as
\begin{equation}\label{eq3.5}
\left(1-\mu\alpha^{2}\cos\theta \right)\frac{\rd^{2}\theta}{\rd s^{2}} + \mu\alpha^{2}\sin\theta\bigl( \frac{\rd\theta}{\rd s} \bigr)^{2} = - \alpha^{2}\sin\theta.
\end{equation}
If we will define that
\begin{equation}\label{eq3.6a}
P(\theta) = 1 - \mu\alpha^{2}\cos\theta=1 - \mu\alpha^{2} + 2\mu\alpha^{2}\sin^{2}\frac{\theta}{2}
\end{equation}
then (\ref{eq3.5}) can be rewritten as
\begin{equation}\label{eq3.6}
P(\theta)\frac{\rd^{2}\theta}{\rd s^{2}} + P'(\theta)\bigl(\frac{\rd\theta}{\rd s} \bigr)^{2} =-\frac{1}{\mu}P'(\theta).
\end{equation}
The above nonlinear second-order ordinary differential equation on function $\theta(s)$ can be once integrated when we multiply the left- and right-hand sides of (\ref{eq3.6}) by the term $2P(\theta)(\rd\theta/\rd s)$. Then we will obtain that
\begin{equation}\label{eq3.7}
\frac{\rd}{\rd s}\bigl(P^{2}(\theta)\bigl(\frac{\rd\theta}{\rd s}\bigr)^{2}\bigr)=-\frac{1}{\mu}\frac{\rd}{\rd s}\left(P^{2}(\theta)\right)
\end{equation}
can be integrated as
\begin{equation}\label{eq3.8}
\bigl(1 + \mu\bigl( \frac{d\theta}{ds} \bigr)^{2} \bigr)P^{2}(\theta)=D^{2} > 0
\end{equation}
from where we obtain that
\begin{equation}\label{eq3.9a}
\frac{\rd\theta}{\rd s}=\pm\sqrt{\frac{1}{\mu}\bigl(\frac{D^{2}}{P^{2}(\theta)} - 1 \bigr)}=\pm\frac{K}{\sqrt{\mu}}\frac{\sqrt{\bigl(1-k_1\sin^{2}\frac{\theta}{2}\bigr)\bigl(1-k_2\sin^{2}\frac{\theta}{2}\bigr)}}{1-k_3\sin^{2}\frac{\theta}{2}}
\end{equation}
where \(D\) is the first integration constant and
\begin{equation}\label{eq3.9b}
k_1=\frac{2\mu\alpha^{2}}{\mu\alpha^{2}-1+D},\ \ k_2=\frac{2\mu\alpha^{2}}{\mu\alpha^{2}-1-D},\ \
k_3=\frac{2\mu\alpha^{2}}{\mu\alpha^{2}-1},\ \ K=\frac{k_3}{\sqrt{k_1k_2}}\cdot
\end{equation}

Let us notice the interesting fact that whereas (\ref{eq3.5}) can be quite easily reduced to (\ref{eq2.5}) when the nonlocality parameter \(\mu\) approaches 0, the connection between (\ref{eq3.9a}) and (\ref{eq2.7}) is not so obvious (because of the division by 0).

Substituting (\ref{eq3.9a}) into (\ref{eq3.6}) we can also obtain the concise expression for the second derivative of \(\theta(s)\), i.e.,
\begin{equation}\label{eq3.10}
\frac{\rd^{2}\theta}{\rd s^{2}}=-\frac{D^{2}}{\mu}\frac{P'(\theta)}{P^{3}(\theta)} = \frac{D^{2}\alpha^{2}\sin\theta}{\left(\mu\alpha^{2}\cos\theta-1\right)^{3}}\cdot
\end{equation}

If we separate the variables in (\ref{eq3.9a}) and integrate the obtained expressions, then
\begin{equation}\label{eq3.11}
s(\theta)=\pm\frac{\sqrt{\mu}}{K}\int
\frac{\bigl(1-k_3\sin^{2}\frac{\theta}{2}\bigr)\rd\theta}{\sqrt{\bigl(1-k_1\sin^{2}\frac{\theta}{2}\bigr)\bigl(1-k_2\sin^{2}\frac{\theta}{2}\bigr)}}
\end{equation}
which leads us also to the corresponding expressions for the coordinates of the deformed elastica, i.e.,
\begin{eqnarray}
x(\theta)&=&\int\cos\theta\;\rd s(\theta)= s(\theta) \mp 2\frac{\sqrt{\mu}}{K}\int
\frac{\bigl(1-k_3\sin^{2}\frac{\theta}{2}\bigr)\sin^{2}\frac{\theta}{2}\;\rd\theta}{\sqrt{\bigl(1-k_1\sin^{2}\frac{\theta}{2}\bigr)\bigl(1-k_2\sin^{2}\frac{\theta}{2}\bigr)}}\nonumber\\
\label{eq3.12-13}\\
y(\theta)&=&\int\sin\theta\;\rd s(\theta) = \pm 4\frac{\sqrt{\mu}}{K}\int
\frac{\bigl(1-k_3\sin^{2}\frac{\theta}{2}\bigr)\sin\frac{\theta}{2}\;\rd\bigl(\sin\frac{\theta}{2}\bigr)}{\sqrt{\bigl(1-k_1\sin^{2}\frac{\theta}{2}\bigr)\bigl(1-k_2\sin^{2}\frac{\theta}{2}\bigr)}}\cdot\nonumber
\end{eqnarray}

Let us notice that the second expression in (\ref{eq3.12-13}) can be quite easily integrated, i.e., we obtain that
\begin{equation}\label{eq3.14}
y(\theta)=y_{0}\mp 2\sqrt{\frac{\mu}{k_1k_2}}\sqrt{\bigl(1-k_1\sin^{2}\frac{\theta}{2}\bigr)\bigl(1-k_2\sin^{2}\frac{\theta}{2}\bigr)}
\end{equation}
where \(y_{0}\) is the integration constant.

\section{Implicit parametrization through incomplete elliptic integrals}\label{sec:4}

Using the substitution \(\tan(\theta/2)=t\), we obtain that
\begin{equation}\label{eq4.1}
\sec^2\frac{\theta}{2}=1+t^2,\quad \sin\theta = \frac{2t}{1 + t^{2}},\quad \cos\theta = \frac{1 - t^{2}}{1 + t^{2}},\quad \rd\theta = \frac{2\;\rd t}{1 + t^{2}}
\end{equation}
therefore, the expressions (\ref{eq3.11}) and the first expression in (\ref{eq3.12-13}) can be rewritten as
\begin{eqnarray}
s(t)&=&\pm\frac{2\sqrt{\mu}}{KK^{\prime}}\int\frac{Q_{2}(t)\rd t}{\left( 1 + t^{2} \right)\sqrt{P_{4}(t)}}\nonumber\\
\label{eq4.2-3}\\
x(t)&=&\pm\frac{2\sqrt{\mu}}{KK^{\prime}}\int\frac{Q_{4}(t)\rd t}{\left( 1 + t^{2} \right)^{2}\sqrt{P_{4}(t)}}\nonumber
\end{eqnarray}
where the polynomials \(Q_{2}(t)\), \(Q_{4}(t)\), and \(P_{4}(t)\) are defined as
\begin{equation}\label{eq4.4}
Q_{2}(t)=k^{\prime}_3-t^{2},\
Q_{4}(t)=k^{\prime}_3-k_3k^{\prime}_3t^{2}+t^{4},\
P_{4}(t)=\bigl(t^{2}-k^{\prime}_1\bigr)\bigl(t^{2}-k^{\prime}_2\bigr)
\end{equation}
with the coefficients given as
\begin{eqnarray}
k^{\prime}_1=\frac{1}{k_1-1}=\frac{\mu\alpha^{2}-1+D}{\mu\alpha^{2}+1-D},&\quad& k_1k^{\prime}_1=\frac{2\mu\alpha^{2}}{\mu\alpha^{2}+1-D}\nonumber\\
\nonumber\\
k^{\prime}_2=\frac{1}{k_2-1}=\frac{\mu\alpha^{2}-1-D}{\mu\alpha^{2}+1+D},&\quad& k_2k^{\prime}_2=\frac{2\mu\alpha^{2}}{\mu\alpha^{2}+1+D}\nonumber\\
\label{eq4.5a-d}\\
k^{\prime}_3=\frac{1}{k_3-1}=\frac{\mu\alpha^{2}-1}{\mu\alpha^{2}+1},&\quad& k_3k^{\prime}_3=\frac{2\mu\alpha^{2}}{\mu\alpha^{2}+1}\nonumber\\
\nonumber\\
K^{\prime}=\frac{k^{\prime}_3}{\sqrt{k^{\prime}_1k^{\prime}_2}},&\quad& \frac{1}{KK^{\prime}}=
\frac{\mu\alpha^2+1}{\sqrt{(\mu\alpha^2+1)^2-D^2}}\cdot\nonumber
\end{eqnarray}

Then the expressions (\ref{eq4.2-3}) can be integrated as
\begin{eqnarray}
s\left(\theta\right)&=&s_0\pm\frac{2\sqrt{\mu}}{KK^{\prime}}\left(k_3k^{\prime}_3I_2\left(\theta\right)-I_1\left(\theta\right)\right)\nonumber\\
\label{eq4.7a}\\
&=&s_0\pm2\sqrt{\mu}\;\frac{2\mu\alpha^2I_2\left(\theta\right)-(\mu\alpha^2+1)I_1\left(\theta\right)}{\sqrt{(\mu\alpha^2+1)^2-D^2}}\nonumber\\\nonumber\\
x\left(\theta\right)&=&x_0-s\left(\theta\right)+s_0\pm\frac{4\sqrt{\mu}}{KK^{\prime}}\left(k_3k^{\prime}_3I_3\left(\theta\right)-I_2\left(\theta\right)\right)\nonumber\\
\label{eq4.7b}\\
&=&x_0-s\left(\theta\right)+s_0\pm 4\sqrt{\mu}\;\frac{2\mu\alpha^2I_3\left(\theta\right)-(\mu\alpha^2+1)I_2\left(\theta\right)}{\sqrt{(\mu\alpha^2+1)^2-D^2}}\nonumber
\end{eqnarray}
where $s_0$ and $x_0$ are the integration constants and
\begin{eqnarray}
I_1\left(\theta\right)&=&\int\frac{\rd t}{\sqrt{P_{4}(t)}}=\sqrt{k_1-1}\;F\left(\varphi\left(\theta\right),k\right)\nonumber\\
\nonumber\\
I_2\left(\theta\right)&=&\int\frac{\rd t}{\left(1+t^{2}\right)\sqrt{P_{4}(t)}}=
\sqrt{k_1-1}\;\Pi\left(n,\varphi\left(\theta\right),k\right)\nonumber\\
\label{eq4.8a-c}\\
I_3\left(\theta\right)&=&\int\frac{\rd t}{\left(1+t^{2}\right)^{2}\sqrt{P_{4}(t)}}=
\left.-\frac{\partial}{\partial \xi}\int\frac{\rd t}{\left(\xi+t^{2}\right)\sqrt{P_{4}(t)}}\;\right|_{\xi=1}\nonumber\\
\nonumber\\
&=&I_2\left(\theta\right)-\left.\sqrt{k_1-1}\;\frac{\partial}{\partial \xi}\;\Pi\bigl(\frac{n}{\xi},\varphi\left(\theta\right),k\bigr)\right|_{\xi=1}.\nonumber
\end{eqnarray}
In the above expressions we have used the definition of the incomplete elliptic integrals of the third kind, i.e.,
\begin{equation}\label{eq4.8d}
\Pi(n,\varphi,k) = \int_{0}^{\varphi}\frac{\rd\varphi}{\bigl( 1 - n\sin^{2}\varphi \bigr)\sqrt{1 - k^{2}\sin^{2}\varphi}}
\end{equation}
where for our case the Jacobi elliptic amplitude \(\varphi(\theta)\), the elliptic modulus \(k\), and the characteristic \(n\) are given as
\begin{eqnarray}
&&\varphi\left(\theta\right)=\arcsin\bigl(\frac{\tan\frac{\theta}{2}}{\sqrt{k^{\prime}_2}}\bigr)=\arcsin\bigl(\sqrt{\frac{\mu\alpha^2+1+D}{\mu\alpha^2-1-D}}\tan\frac{\theta}{2}\bigr)\nonumber\\
\label{eq4.9a-b}\\
&&k=\sqrt{\frac{k^{\prime}_2}{k^{\prime}_1}}=\sqrt{\frac{\left(\mu\alpha^2-D\right)^2-1}{\left(\mu\alpha^2+D\right)^2-1}},\qquad n=-k^{\prime}_2=\frac{1-\mu\alpha^{2}+D}{1+\mu\alpha^{2}+D}\cdot\nonumber
\end{eqnarray}

\begin{figure}[h!]
\centering
\includegraphics[width=12cm,keepaspectratio]{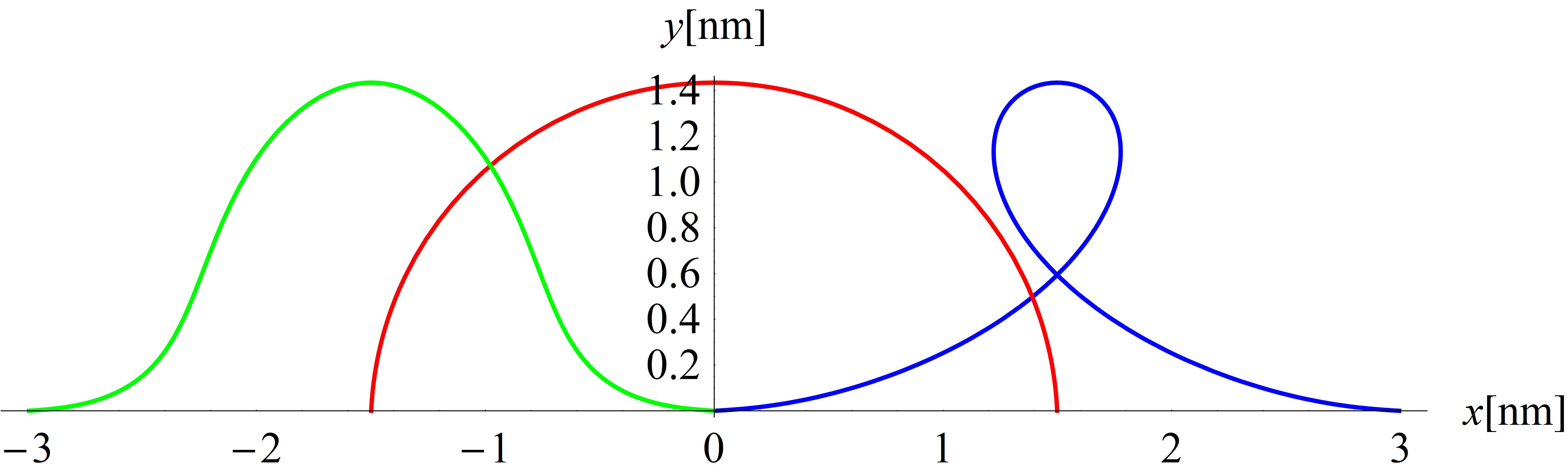}
\caption{Exemplary graphs of the parametric solution $(x\left(\theta\right),y\left(\theta\right))$ defined by (\ref{eq4.13a-b}) and (\ref{eq3.14}) for integration constants $x_0=y_0=0$, $D=0.1$, nonlocality parameter $\mu=2.25$ nm$^2$, force $F=500$ nN, Young modulus $E=1000$ GPa, moment of inertia $I=0.0491$ nm$^{4}$, where green line: $-\pi/2+\varphi(\theta)$, red line: $\varphi(\theta)$, blue line: $\pi/2-\varphi(\theta)$.} \label{fig-2}
\end{figure}
In order to rewrite the last expression in (\ref{eq4.8a-c}), we use the formula for the derivative of the incomplete elliptic integral of the third kind with respect to its characteristic, i.e.,
\begin{eqnarray}
&&\left.-\frac{\partial}{\partial \xi}\;\Pi\bigl(\frac{n}{\xi},\varphi,k\bigr)\right|_{\xi=1}=\frac{n}{2(n-1)(k^2-n)}\bigl(\frac{n\sin2\varphi\sqrt{1-k^2\sin^2\varphi}}{2\bigl(n\sin^2\varphi-1\bigr)}\bigr.\nonumber\\
\label{eq4.10}\\
&&\bigl.+\frac{k^2-n}{n}\;F\left(\varphi,k\right)+E\left(\varphi,k\right)+\frac{n^2-k^2}{n}\Pi\left(n,\varphi,k\right)\bigr)\nonumber
\end{eqnarray}
and the connections between formulas (\ref{eq4.5a-d}) and (\ref{eq4.9a-b}), i.e.,
\begin{eqnarray}
2(n-1)&=&-2(k^{\prime}_{2}+1)=-2k_2k^{\prime}_{2}\nonumber\\
k^2-n&=&\frac{k^{\prime}_{2}}{k^{\prime}_{1}}+k^{\prime}_{2}=k_1k^{\prime}_{2}\label{eq4.11.1-3}\\
n^2-k^2&=&k^{\prime 2}_{2}-\frac{k^{\prime}_{2}}{k^{\prime}_{1}}=\frac{k^{\prime}_{2}}{k^{\prime}_{1}}(k^{\prime}_{1}k^{\prime}_{2}-1)\nonumber
\end{eqnarray}
so that
\begin{eqnarray}
\frac{n}{2(n-1)(k^2-n)}&=&\frac{1}{2k_1k_2k^{\prime}_{2}}\nonumber\\
1+\frac{n^2-k^2}{2(n-1)(k^2-n)}&=&\frac{(2k_1k_2-1)k^{\prime}_{1}k^{\prime}_{2}+1}{2k_1k_2k^{\prime}_{1}k^{\prime}_{2}}\label{eq4.11.4-6}\\
&=&\frac{3+2(k^{\prime}_1+k^{\prime}_2)+k^{\prime}_1k^{\prime}_2}{2k_1k_2k^{\prime}_{1}k^{\prime}_{2}}\nonumber
\end{eqnarray}
then we obtain that
\begin{eqnarray}
I_3\left(\theta\right)&=&-\frac{I_1\left(\theta\right)}{2k_2k^{\prime}_2}+\frac{3+2\left(k^{\prime}_1+k^{\prime}_2\right)+k^{\prime}_1k^{\prime}_2}{2k_1k_2k^{\prime}_1k^{\prime}_2}\;I_2\left(\theta\right)+\frac{\sqrt{k_1-1}}{2k_1k_2k^{\prime}_2}\;E(\varphi\left(\theta\right),k)\nonumber\\
\label{eq4.11}\\
&+&\frac{\tan\frac{\theta}{2}}{2k_1k_2\sqrt{k^{\prime}_1k^{\prime}_2}}\;\sqrt{\bigl(1-k_1\sin^{2}\frac{\theta}{2}\bigr)\bigl(1+k_2\sin^{2}\frac{\theta}{2}\bigr)}.\nonumber
\end{eqnarray}
Taking into account (\ref{eq3.14}), we can rewrite the last term in (\ref{eq4.11}) as
\begin{equation}\label{eq4.11a}
\frac{\tan\frac{\theta}{2}}{\sqrt{k_1k_2k^{\prime}_1k^{\prime}_2}}
\bigl(\mp\frac{y\left(\theta\right)-y_0}{4\sqrt{\mu}}\bigr)
\end{equation}
therefore, after substitution of (\ref{eq4.11}) and (\ref{eq4.11a}) into (\ref{eq4.7b}) we obtain that
\begin{eqnarray}
x\left(\theta\right)&=&x_0-\left(y(\theta)-y_0\right)\tan\frac{\theta}{2}-s\left(\theta\right)+s_0\pm 2\sqrt{\mu}\; \bigl(\frac{E(\varphi\left(\theta\right),k)}{\sqrt{k_1k_2k^{\prime}_2}}\bigr.\nonumber\\
\label{eq4.12}\\
&-&\sqrt{\frac{k_1k^{\prime}_1}{k_2k^{\prime}_2}}\;I_1\left(\theta\right)+\bigl.\bigl(\frac{3+2\left(k^{\prime}_1+k^{\prime}_2\right)+k^{\prime}_1k^{\prime}_2}{\sqrt{k_1k_2k^{\prime}_1k^{\prime}_2}}-\frac{2}{KK^{\prime}}\bigr)I_2\left(\theta\right)\bigr).\nonumber
\end{eqnarray}
Finally, substituting the first two expressions in (\ref{eq4.8a-c}) into (\ref{eq4.7a}) and (\ref{eq4.12}) and simplifying them, we obtain the implicit parametrization of $s(\theta)$ and $x(\theta)$ through incomplete elliptic integrals  (see Fig.~\ref{fig-2} for the exemplary graphs of the obtained general solution), i.e.,
\begin{eqnarray}
s\left(\theta\right)&=&s_0\pm 2\sqrt{\mu}\;\frac{2\mu\alpha^2\;\Pi\left(n,\varphi\left(\theta\right),k\right)-\bigl(1+\mu\alpha^2\bigr)F(\varphi\left(\theta\right),k)}{\sqrt{(\mu\alpha^2+D)^2-1}}\nonumber\\
x\left(\theta\right)&=&x_0-\left(y(\theta)-y_0\right)\tan\frac{\theta}{2}\label{eq4.13a-b}\\
&& \pm \frac{\sqrt{(\mu\alpha^2+D)^2-1}}{\sqrt{\mu}\alpha^2}E(\varphi\left(\theta\right),k)\mp\frac{2\sqrt{\mu}D\;F(\varphi\left(\theta\right),k)}{\sqrt{(\mu\alpha^2+D)^2-1}}\nonumber
\end{eqnarray}
where $y(\theta)$ is given by the expression (\ref{eq3.14}).

\section{Final remarks}\label{sec:5}

As further research, we plan to apply the chosen boundary conditions (clamped, simply supported, etc.) to the obtained nonlocal general solutions (\ref{eq4.13a-b}) and (\ref{eq3.14}) of the deformed beam in order to compare the local and nonlocal versions of the Euler's elastica for the same kinds of the boundary value problems (BVPs).

\end{document}